\documentclass[12pt]{article}
\usepackage{amssymb,amsmath,amscd,amsthm}
\usepackage[cp1251]{inputenc}
\usepackage[T2A]{fontenc}
\usepackage{pstricks}
\usepackage{graphicx,psfrag}

\usepackage[english]{babel}

\newtheorem{theorem}{Theorem}[section]

\newtheorem{lemma}[theorem]{Lemma}

\date{}

\begin{document}

\title{Bounds on positive interior transmission eigenvalues}

\author{ E.Lakshtanov\thanks{Department of Mathematics, Aveiro University, Aveiro 3810, Portugal.   This work was supported by {\it FEDER} funds through {\it COMPETE}--Operational Programme Factors of Competitiveness (``Programa Operacional Factores de Competitividade'') and by Portuguese funds through the {\it Center for Research and Development in Mathematics and Applications} (University of Aveiro) and the Portuguese Foundation for Science and Technology (``FCT--Fund\c{c}\~{a}o para a Ci\^{e}ncia e a Tecnologia''), within project PEst-C/MAT/UI4106/2011 with COMPETE number FCOMP-01-0124-FEDER-022690, and by the FCT research project
PTDC/MAT/113470/2009 (lakshtanov@rambler.ru).} \and
B.Vainberg\thanks{Department
of Mathematics and Statistics, University of North Carolina,
Charlotte, NC 28223, USA. The work was partially supported   by the NSF grant DMS-1008132 (brvainbe@uncc.edu).}}

\maketitle

\begin{abstract}
The paper contains lower bounds on the counting function of the positive eigenvalues of the interior transmission problem when the latter is elliptic. In particular, these bounds justify the existence of an infinite set of interior transmission eigenvalues and provide asymptotic estimates from above on the counting function for the large values of the wave number. They also lead to certain important upper estimates on the first few interior transmission eigenvalues. We consider the classical transmission problem as well as the case when the inhomogeneous medium contains an obstacle.
\end{abstract}


\section{Introduction.}
Interior transmission eigenvalues (ITE-s) were introduced in the middle of 1980s  and soon became a classical object in the scattering theory, see, e.g., a recent review \cite{HadCak}. Their importance is based on the relation of ITEs to the far-field operator: if real $\lambda=k^2$ is not an ITE, then the far-field operator with the wave number $k$
 is injective and has a dense range. In particular,  when the linear sampling method (widely used in the inverse scattering theory) is applied for recovery of the support of the inhomogeneity in the medium, one needs to know  that the far-field operator has a dense range, i.e., $\lambda=k^2$ is not an ITE.  For this and other applications, it is important to know not only the fact of the discreteness of the ITEs but also their distribution. Note that ITEs can be measured, and this opens an opportunity to use ITEs for the recovery of the properties of the scatterer (eg \cite[Th.3.2]{HadCak}).

Let us recall the definition of ITEs.
The values of $\lambda   \in \mathbb C$ for which the homogeneous problem
\begin{equation}\label{Anone0}
-\Delta u - \lambda u =0, \quad x \in \mathcal O, \quad u\in H^2(\mathcal O),
\end{equation}
\begin{equation}\label{Anone}
-\nabla A \nabla v - \lambda   n(x)v =0, \quad x \in \mathcal O, \quad v\in H^2(\mathcal O),
\end{equation}
\begin{equation}\label{Antwo}
\begin{array}{l}
u-v=0, \quad x \in \partial \mathcal O, \\
\frac{\partial u}{\partial \nu} - \frac{\partial v}{\partial \nu_A}=0, \quad x \in \partial \mathcal O,
\end{array}
\end{equation}
has a non-trivial solution are called the \textit{interior transmission eigenvalues}.
Here $\mathcal O\subset R^d$ is a bounded domain with a $C^\infty$-boundary, $H^{2}(\mathcal O), ~H^{s}(\partial \mathcal O)$ are the Sobolev spaces, $A(x),~x\in \overline{\mathcal O}$, is a smooth symmetric elliptic ($A=A^t>0$) matrix with real-valued entries, $n(x)$ is   a smooth function, $\nu$ is the outward unit normal vector, and the co-normal derivative is defined as follows
$$
\frac{\partial } {\partial \nu_A}v =\nu \cdot A \nabla v.
$$
Problem (\ref{Anone0})-(\ref{Antwo}) appears naturally when the scattering of plane waves is considered, and the inhomogeneity in $R^d$ is located in $\mathcal O$ and described by an anisotropy matrix $A$ and the refraction index $n$.  We will be mostly concerned with the case $d=2,3$, but all the results below can be automatically carried over to any dimension $d$.

We will also consider the case when $\mathcal O$ contains a compact obstacle $\mathcal V \subset \mathcal O$, $\partial V \in C^\infty$.
In this case, equation (\ref{Anone}) is replaced by
\begin{equation}\label{AnoneB}
-\nabla A \nabla v - \lambda   n(x)v =0, \quad x \in \mathcal O \backslash V, \quad v\in H^2(\mathcal O \backslash V);~~~
v(x)=0, \quad x \in \partial V,
\end{equation}
while equation (\ref{Anone0}) remains valid in $\mathcal O$. For simplicity of notations, we will consider problem (\ref{Anone0})-(\ref{Antwo}) as a particular case of (\ref{Anone0}),(\ref{AnoneB}),(\ref{Antwo}) with $\mathcal V = \emptyset$.

Denote the set of real {\it non-negative}  ITEs with their multiplicity taken into account by $\{\lambda^T_i\}$. Similarly, denote the set of positive eigenvalues of the Dirichlet problem for $-\Delta$ in $\mathcal O$ by $\{\lambda_i\}$, and the set of positive $\lambda>0$ for which equation (\ref{AnoneB}) in $\mathcal O \backslash V$ with the Dirichlet boundary conditions at the boundary $\partial\left(\mathcal O \backslash   V\right)$ has a nontrivial solution by $\{ \lambda^A_i\}$.
The corresponding counting functions will be denoted by
\begin{equation}\label{countdef}
N_T(\lambda)=    \#\{i:\lambda^T_i \leq \lambda\} , \quad N(\lambda)=     \#\{i:\lambda_i \leq \lambda\}, \quad N_A(\lambda)=     \#\{i:\lambda^A_i \leq \lambda\}.
\end{equation}

Let us stress that problem (\ref{Anone0}),(\ref{Antwo}),(\ref{AnoneB}) is not symmetric, and the existence of the real eigenvalues can not be established by soft arguments. Note, that even in the case when the set of positive ITEs is countable, they could be distributed so sparsely or so densely, that from the practical point of view, the situation would be the same as in the case when the set of ITE is finite or not discrete, respectively. Thus it is important to know conditions for the set  $\{\lambda^T_i\}$ to be discrete (counter examples can be found in \cite{lakvain5}), to be infinite, as well as to know the asymptotic behavior of $N_T(\lambda)$ as $\lambda\to\infty$.

It is known (due to F.Cakoni, D.Gintides, H.Haddar, A.Kirsch, e.g., \cite[th.4.8]{HadCak},\cite{Dr},\cite{CK}) that the set $\{\lambda^T_i\}$ of non-negative ITEs is infinite if $\mathcal V = \emptyset$, and the matrix   $(1-n)(A-I)$ is positive-definite for all $x\in \overline{\mathcal O}$.
When $A=I$, some estimates on $N(\lambda)$ for ITEs can be found in   \cite{ss},\cite{tsm}.

The case $\mathcal V \neq \emptyset$ was studied in \cite{chobst}.   The existence of infinitely many real ITEs was proved for the isotropic media $A=I$ if $n<1$ everywhere.   The existence of at least one real eigenvalue was shown if $A(x)>I,~ n \neq 1,~ x \in \overline{\mathcal O}$, and the obstacle is small enough. The authors of the latter paper noted that the case $A<I, \mathcal V \neq \emptyset$ remains unstudied.

The discreteness of the spectrum and the upper estimate on $N_T(\lambda) $:
$$
N_T(\lambda) \leq \lambda^{d/2} \frac{\omega_d}{(2\pi)^d} \int_{\mathcal O} \left (1+\frac{n^{d/2}(x)}{(det A(x))^{1/2}} \right ) dx + o(\lambda^{d/2}), \quad \lambda \rightarrow \infty,
$$
where $\omega_d$ is the volume of the unit
ball in $\mathbb R^d$, were established in \cite{lakvain5}\footnote{The result from \cite{BK} was used essentially} (in the case of $\mathcal V = \emptyset$) under minimal assumptions on $A$ and $n$ imposed only at the boundary of the domain (even the sign-definiteness of the matrix $A-I$ at the boundary was not required). Since these assumptions and results are relevant to the present paper, they will be reviewed in the Appendix.  It will be also explained in the Appendix that the results of \cite{lakvain5} mentioned above  are valid in the case of the presence of the obstacle $\mathcal V $ and their proofs remain the same. The lower estimate on $N_T(\lambda)$ under the same minimal assumptions on $A$ and $n$ will also be proved below.

This paper contains the following estimate on $N_T(\lambda)$ from below through the counting functions $N(\lambda),N_A(\lambda)$:
\begin{theorem}\label{lemmamain1} Let
 the interior transmission problem be elliptic   (i.e., the assumptions of Theorem \ref{a3} hold) and let the set of ITEs be discrete (for example, the assumptions of either Theorem \ref{cond1} or Theorem \ref{cond2} hold). Then there exist constants $\sigma=\pm 1$ and $n^-(0)\geq 0$ such that
\begin{equation}\label{mainIneq}
N_T(\lambda) \geq \sigma(N(\lambda)-N_A(\lambda))-n^-(0), \quad \lambda >0.
\end{equation}
\end{theorem}
\textbf{Remark.} The constants   $\sigma$ and $n^-(0)$ will be defined in the next section. In particular,  $\sigma=sgn(I-A)$ if $I-A$ is sign definite at the boundary $\partial \mathcal O$.

The following result is one of the important direct consequences of Theorem \ref{lemmamain1} and the well-known (e.g., \cite[Th. 1.2.1]{safvas}) Weyl formula for $N(\lambda)$ and $N_A(\lambda)$.
\begin{theorem}\label{th12} Let
 the interior transmission problem be elliptic   (i.e. assumption of the Theorem \ref{a3} hold) and let the set of ITEs be discrete (for example, assumptions of either Theorem \ref{cond1} or Theorem \ref{cond2} hold).
If
$$
\gamma := \sigma \left ( Vol(\mathcal O) - \int_{\mathcal O \backslash \mathcal V} \frac{n^{d/2}(x)dx}{(det A(x))^{1/2}} \right )>0,
$$
then the set of positive ITEs is infinite, and moreover,
$$
N_T(\lambda)   \geq \frac{\omega_d}{(2\pi)^d} \gamma   \lambda^{d/2} + O(\lambda^{(d-1)/2}), \quad \lambda \rightarrow \infty.
$$
\end{theorem}
The next theorem provides a condition when the constant $n^-(0)$ in (\ref{mainIneq}) can be omitted.
\begin{theorem}\label{lemmamain2}
Let the matrix $I\!-\!A$ be sign-definite in the whole $\overline{\mathcal O }$,  and let the set of ITEs be discrete (for example, the assumptions of either Theorem \ref{cond1} or Theorem \ref{cond2} hold). Suppose that the set $\mathcal V$ is empty.
Then $n^-(0)=0$, and therefore
$$
N_T(\lambda) \geq \sigma(N(\lambda)-N_A(\lambda)), \quad \lambda >0.
$$
The same statement is valid in the presence of an obstacle  $\mathcal V$ if $A>I$ on $\overline{\mathcal O} \backslash \mathcal V$.
\end{theorem}

Consider a particular case of problem (\ref{Anone0}),(\ref{Antwo}),(\ref{AnoneB}) when $A=aI$, where $ a>0$ and the refruction index $ n>0$ are constant, and $ an \neq 1, a \neq 1$. Then Theorem \ref{th12} takes the form: if
\begin{equation}\label{coef1}
\gamma=\mbox{sgn}(1-a) \left (    Vol(\mathcal O)  - \left (\frac{n}{a}\right)^{d/2} Vol(\mathcal O \backslash \mathcal V ) \right )>0,
\end{equation}
then the set of positive ITEs is infinite and
$$
N_T(\lambda) \geq  \frac{\omega_d}{(2\pi)^d} \gamma \lambda^{d/2} + O(\lambda^{(d-1)/2}), \quad \lambda \rightarrow \infty.
$$

Note that $\lambda^T_0=0$ is an ITE in the simple case of problem  (\ref{Anone0})-(\ref{Antwo}) described above.  The  Faber-Krahn type inequality \cite{HadCak} provides the following bound from below on the first positive ITE (e.g., \cite[Th.4.9]{HadCak}):
$$
\lambda^T_1 \geq \frac{a}{n}\lambda_1, \quad a<1; \quad \quad \lambda^T_1 \geq \lambda_1, \quad a>1.
$$
Inequality (\ref{mainIneq}) allows one to estimate the first few positive ITEs $\lambda^{T}_i$ from above through $\{\lambda_i\}$ and $\{ \lambda^A_i\}$. We will demonstrate this by providing an estimate on $\lambda^{T}_1$ in the same simple case of constant $A$ and $n$.
Inequality (\ref{mainIneq}) implies that
 the first strictly positive ITE (the second non-negative) $\lambda^{T}_1$ can not exceed a root of the equation
$$
\sigma(N(\lambda)-N_A(\lambda))-n^-(0)=2.
$$
This will be used to show (section 2) that the following theorem is valid
\begin{theorem}\label{ThEstFK}
Let $A=a I$, where $a>0$ is a constant, and let $ n>0$ be constant. Assume also that $an \neq 1, a \neq 1$. Then the following inequalities are valid.
\begin{itemize}
\item If $a<1$ and $n \lambda_2 < a \lambda_1$, then
$$
\lambda^{T}_1 \leq \lambda_2.
$$
\item If $a>1$ and $a \lambda_2 < n \lambda_1$, then
$$
\lambda^T_1 \leq \frac{a}{n}\lambda_2.
$$
\end{itemize}
\end{theorem}
{\bf Remark.}
 If $\frac{\lambda_2}{\lambda_1} < \left ( \frac{a}{n} \right )^\sigma$ is violated (i.e., the distance between $\lambda_1$ and $\lambda_2$ is not small enough), then similar estimates on $\lambda^{T}_1$ can be obtained using next eigenvalues of the Dirichlet problem. For example, if $\sigma=1$
and $\frac{\lambda_3}{\lambda_2} <  \frac{a}{n} $, then
$\lambda_1^T<\lambda_3.$  In the same way one can obtain the inequalities for higher ITEs $\lambda^T_i, i \geq 2$. For example, if $\sigma=-1$ and $\frac{\lambda_3}{\lambda_1}<\frac{n}{a}$, then $\lambda_2^T<\frac{a}{n}\lambda_3.$

\textbf{Comments and acknowledgments.}
This work was inspired by several papers. One of them is Eckmann and Pillet paper \cite{EP2} on the relations between $N(\lambda)$ and the total phase of the scattering matrix for the problem of scattering by an obstacle. Another important impulse was made by Safarov's paper \cite{saf}) on the difference between the counting functions for the Dirichlet and Neumann problems. Our method can be considered as a generalization of Friedlander's idea \cite{fried}, who considered the Dirichlet-to-Neuman map $F(\lambda)$ for the Helmholtz equation and noted that the passing of   $\lambda$ through $\lambda_i$ implies the transition of the eigenvalues of $F(\lambda)$ from $\mathbb R^-$ to $\mathbb R^+$, and never backwards. The authors are grateful to H.Haddar who attracted their attention to the specific case of the problem with an obstacle in the medium.

Perhaps our approach can be applied to more general non-selfadjoint systems and lead to an estimate on the counting function of the system through the counting functions of the individual equations.

\section{Proof of main results}\label{SectionTheor}

\subsection{Proof of Theorem \ref{lemmamain1}.}
{\it Description of ITEs through the Dirichlet-to-Neumann maps.} The Dirichlet problem is uniquely solvable for equations (\ref{Anone0}), (\ref{Anone}) or (\ref{AnoneB}) when parameter $ \lambda $ is not an eigenvalue of the corresponding Dirichlet problem. Hence, the Dirichlet-to-Neumann map $F(\lambda)$ is defined for equation (\ref{Anone0}) if $ \lambda \notin \{ \lambda_i\}$ and   the Dirichlet-to-co-normal derivative map $F_A(\lambda)=F(A,n,\mathcal V,\lambda)$ is defined for equations   (\ref{Anone}) or (\ref{AnoneB}) when $ \lambda \notin \{ \lambda^A_i\}$. The operators
\begin{equation}\label{ffa}
F(\lambda), F_A(\lambda):H^{3/2}(\partial\mathcal O)\to H^{1/2}(\partial\mathcal O)
\end{equation}
are elliptic pseudo-differential operators of the first order which depend analytically on $\lambda$ with poles at the sets $ \{ \lambda_i\}$ and $ \{ \lambda^A_i\}$, respectively. Operators $F,F_A$ can be expressed through resolvents $R_\lambda, R_{\lambda, A}$ of the Dirichlet problem (with zero boundary condition) for operators $-\Delta, -\nabla A \nabla$, respectively. For example,
$$
F\phi=\frac{\partial}{\partial\nu}[T\phi-R_\lambda((\Delta  + \lambda) T\phi)]|_{\partial\mathcal O},
$$
where
$T:H^{3/2}(\partial \mathcal O)\to H^2(\mathcal O)$ is a bounded extension operator. Since the resolvent of a self-adjoint operator may have poles of at most first order, operators $F,F_A$ also have poles of at most first order. Obviously, the residues at the poles of $F,F_A$ are finite dimensional symmetric operators. Let $\lambda=\lambda_0\in \{ \lambda_i\}\bigcup\{ \lambda^A_i\}$ be a pole of $F(\lambda)-F_A(\lambda)$, i.e.,
\begin{equation}\label{defRF}
F(\lambda)-F_A(\lambda)=\frac{P}{\lambda-\lambda_0}+Q(\lambda),
\end{equation}
where the operator $Q$ is analytic in $\lambda$ in a neighborhood of $\lambda_0$. The term 'kernel of $F(\lambda)-F_A(\lambda)$'  will be used not only when the operator is analytic, but also when $\lambda=\lambda_0$ is a pole of the operator. In the latter case, the kernel will be understood as the set of all $f\in H^{3/2}(\partial\mathcal O)$ such that $Pf=Q(\lambda_0)f=0.$

The following statement is an immediate consequence of the definition of ITEs.
\begin{lemma}\label{ites}
A point $\lambda=\lambda_0$ is an ITE if and only if the operator $F(\lambda_0) - F_A(\lambda_0)$ has a non-trivial kernel or the following two conditions hold:

1) $\lambda=\lambda_0$ is an eigenvalue of the Dirichlet problem for $-\Delta$ and for equation (\ref{AnoneB}), i.e.,   $\lambda=\lambda_0$ is a pole for both $F(\lambda)$ and $F_A(\lambda)$.

2) The ranges of the residues of operators $F(\lambda)$ and $F_A(\lambda)$ at the pole $\lambda=\lambda_0$ have a non trivial intersection.

Moreover, the multiplicity of the interior transmission eigenvalue $\lambda=\lambda_0$ in all cases is equal to $m_1+m_2$, where $m_1$ is the dimension of the kernel of the operator $F(\lambda)-F_A(\lambda)$, and   $m_2$ is the   dimension of the intersection of the ranges of the residues  at the pole $\lambda=\lambda_0$ ($m_2=0$ if $\lambda=\lambda_0$ is not a pole).
\end{lemma}
If $\lambda=\lambda_0$ satisfies the latter two conditions, we will call it a singular spectral point. Thus, singular spectral points belong to the intersection of three spectral sets: $ \{\lambda^{T}_i\},~\{\lambda_i\}$ and $\{ \lambda^A_i\}$.

It will be used several times below that operators  $F_A(\lambda)$ and $F(\lambda)$ are symmetric for real $\lambda$. The latter follows immediately from the Green formulas for equations (\ref{Anone0}) and (\ref{AnoneB}).

{\it Definition of $\sigma$.} Principal symbols $p(x^0,\tau),p_A(x^0,\tau)$ (where $x^0\in \partial \mathcal O ,~\tau\in R^{d-1}$) of elliptic operators (\ref{ffa}) can be easily written down using the following procedure, see, e.g., \cite[2.7]{KS} . Let $d=3$. For an arbitrary point $x^0\in \partial \mathcal O $, we choose local coordinates $y=C(x-x^0)$, where the $y_3$-axis is directed along the outward normal $\nu$ and $C=C_{x_0}$ is the orthogonal transfer matrix. Let $a_{i,j}$ be the entries of the matrix $\widetilde{A}(x^0)=CA(x^0)C^*$. Then
\begin{equation}\label{symbols}
p=|\tau|, \quad
p_A=\sqrt{a_{3,3}(\sum_{i,j=1}^2a_{i,j}\tau_i\tau_j)-(a_{1,3}\tau_1+a_{2,3}\tau_2)^2},~~\tau\in R^2.
\end{equation}
These symbols are obtained by evaluating the Dirichlet to Neumann maps for the  equations in the half space given in (\ref{Anone1}), the first equation is used to find $p$, and the second to find $p_A$.

 Assumption of Theorem \ref{lemmamain1} on the ellipticity of the problem (\ref{Anone0}),(\ref{AnoneB}),(\ref{Antwo}) implies (see Remarks 1 and 3 in the Appendix) that the pseudo-differential operator $\sigma(F_A(\lambda)-F(\lambda))$ is elliptic, i.e., its principal symbol $p_A-p$ does not vanish for all $x^0 \in \partial \mathcal O,~\tau\neq 0$. Let us introduce $\sigma=\textrm{sgn}(p-p_A),~\tau \neq 0$. An equivalent definition of $\sigma$ is given in Remark 2 in the Appendix. Note that if the matrix $A-I$ is sign-definite on the boundary $\partial \mathcal O$ of the domain $\mathcal O$, then the problem (\ref{Anone0}),(\ref{AnoneB}),(\ref{Antwo}) is elliptic, and $\sigma=\textrm{sgn}(I-A),~x\in \mathcal O$, see remark 2 after Assumption 2.4 in \cite{LV4}.

The operator $\sigma(F(\lambda)-F_A(\lambda))$ has a positive principal symbol, and therefore (see \cite[Cor. 9.3]{shubin}) it is bounded from below when $\lambda$ is not a pole. Obviously, the bound can be chosen uniformly in $\lambda$, i.e., the following statement holds.
\begin{lemma}\label{lemmaShubin}
For each closed interval $\pi$ of the $\lambda$-axis where the operator $\sigma(F(\lambda)-F_A(\lambda))$ is analytic, there exists a constant $C=C(\pi)$ such that
\[
\sigma(F(\lambda)-F_A(\lambda))\geq -C, \quad   \lambda\in\pi.
\]
\end{lemma}

{\it Analytic properties of eigenvalues $\mu_j=\mu_j(\lambda)$ of the operator $\sigma(F(\lambda)-F_A(\lambda))$.}\label{SUBSECTANAL} For each fixed real $\lambda\in \mathbb R$,  which is not a pole of the operator $\sigma(F(\lambda)-F_A(\lambda))$, consider the eigenvalues $\mu_j=\mu_j(\lambda)$ of the operator $\sigma(F(\lambda)-F_A(\lambda))$. We will discuss the properties of the eigenvalues $\mu_j$ here, and their relation to the set $\{\lambda_i^T\}$ of ITEs in the next subsection.


It was mentioned above that $\sigma(F(\lambda)-F_A(\lambda))$ is a symmetric elliptic pseudo-differential operator of the first order.
Hence the spectrum of the operator $\sigma(F(\lambda)-F_A(\lambda)), ~\lambda\in R,$ is discrete and consists of the set of real eigenvalues $\{\mu_j(\lambda)\}$.

\begin{lemma}\label{hardDayLemma}
If the operator $\sigma(F(\lambda_0)-F_A(\lambda_0))$  is analytic in a neighborhood of a point $\lambda=\lambda_0$, then all the eigenvalues $ \mu=\mu_j(\lambda)$ are analytic in this neighborhood.

If  $\lambda=\lambda_0$ is a pole of the operator  $\sigma(F(\lambda_0)-F_A(\lambda_0))$ and $p$ is the rank of the residue $P$ (see (\ref{defRF})), then $p$ eigenvalues  $ \mu=\mu_j(\lambda)$ have a pole at
$\lambda_0$ and all the others are analytic in this neighborhood. Moreover, if
 $\mu_0 \in \mathbb R,f_0 \in H^{3/2}(\partial \mathcal O)$ are an eigenvalue and an eigenfunction of the operator   $(I - \mathcal P ) \sigma(F(\lambda_0)-F_A(\lambda_0))(I - \mathcal P)$, where $\mathcal P$ is the orthogonal projection on $Ker P$, then they can be extended analytically in a neighborhood of $\lambda _0$ as an eigenvalue and an eigenfunction of  $\sigma(F(\lambda)-F_A(\lambda)), \quad \lambda \neq \lambda_0$.
\end{lemma}
{\bf Proof.} The first statement is a well-known property of analytic self adjoint operators (see \cite{shubin}). In order to prove the second property, consider the operator $A(\lambda)=(\lambda-\lambda_0)\sigma(F_A(\lambda)-F(\lambda))$. It is analytic in a neighborhood of $\lambda=\lambda_0$ and has exactly $p$ eigenvalues which do not vanish at $\lambda_0$. Let $D=D_\lambda$ be the $p$-dimensional space spanned by the corresponding eigenfunctions. By using $D$ and its orthogonal complements, one can write the original operator in a block form, where the block which corresponds to $D$ has a pole and the second block is analytic. After that the statement of lemma follows immediately from the general properties of analytic families of self-adjoint operators.
\qed

{\it Relation between the set of ITEs $\lambda_i^T$ and the eigenvalues $\mu_j=\mu_j(\lambda)$}.
Denote by $n^-(\lambda),~ \lambda\notin\{ \lambda_i\}\bigcup\{ \lambda^A_i\}\bigcup\{\lambda^T_i\}$,
the number of the negative eigenvalues $\mu_j(\lambda)$ of the operator $\sigma(F(\lambda)-F_A(\lambda))$. From Lemmas \ref{lemmaShubin} and \ref{hardDayLemma} it follows that this number is finite for each $\lambda$.

Let us evaluate the difference $n^-(\lambda')-n^-(0)$ by moving $\lambda$ from $\lambda=0$ to $\lambda=\lambda'>0$. Since the eigenvalues    $\mu_j(\lambda)$ are meromorphic functions of   $\lambda$, the number of negative eigenvalues $\mu_j(\lambda)<0$ changes only when some of them pass through the 'edges' of the interval $(-\infty,0)$, i.e.,
\begin{equation}\label{n10}
n^-(\lambda')-n^-(0)=n_1(\lambda')+n_2(\lambda'),
\end{equation}
where $n_1(\lambda')$ is the change in $n^-$ due to passing some eigenvalues through $\mu=-\infty$ and  $n_2(\lambda')$ is the change in $n^-$ due to passing some eigenvalues through $\mu=0$.
The annihilation or the birth of $\mu_j(\lambda)$ at $\mu=-\infty$ may occur only when $\lambda$ passes through a pole $\lambda=\lambda_0$ of the operator $\sigma(F(\lambda)-F_A(\lambda))$.

 Let us denote by $\delta n_1(\lambda_0)$ the jump of $n_1$ at a pole $\lambda=\lambda_0$  due to  passing of $\mu_j$ through infinity. The following lemma will be proved below.
\begin{lemma}\label{ms}
The following relation holds for every pole $\lambda=\lambda_0>0$ of the operator $\sigma(F(\lambda)-F_A(\lambda))$:
\begin{equation}\label{mainlemma}
|\delta n_1(\lambda_0)-\sigma(m_A-m_0)|\leq m.
\end{equation}
Here $m_A$ and $m_0$ are the dimensions of the residues of operators  $F_A(\lambda)$ and $F(\lambda)$, respectively, at the pole, and $m$ is the dimension of the intersection of the ranges of the residues.

In particular, if $\lambda_0\notin \{\lambda^T_i\}$, then $m=0$ and
\begin{equation}\label{mainlemma1}
\delta n_1(\lambda_0)=\sigma(m_A-m_0).
\end{equation}
\end{lemma}

By the summation of inequalities (\ref{mainlemma}) over all poles $\lambda_0$ on the interval $(0,\lambda)$, we obtain the following relation: 
\begin{equation}\label{n1}
|n_1(\lambda)-\sigma(N(\lambda)-N_A(\lambda))|\leq R(\lambda),
\end{equation}
where $R(\lambda)$ is the counting function of the singular interior transmission eigenvalues. Their multiplicities are defined exclusively by the dimensions $m=m(\lambda_0)$ of the intersection of the ranges of the residues. (Recall, the multiplicity of an ITE $\lambda_0\in  \{\lambda_i^T\}$ is the sum of $m$ and the dimension of the kernel of the operator $\sigma(F(\lambda_0)-F_A(\lambda_0))$, see Lemma \ref{ites}.) From (\ref{n10}) and (\ref{n1}) it follows that 
\begin {equation}\label{nnn}
n^-(\lambda)-n^-(0)+\sigma(N_A(\lambda)-N(\lambda))\leq R(\lambda)+n_2(\lambda).
\end{equation}

Consider now all the interior transmission eigenvalues $  \{\lambda_s^T\}$ which were not counted by the function $R(\lambda)$. In a neighborhood of each of these transmission eigenvalues, there exists an analytic in $\lambda $ eigenvalue $\mu_i(\lambda)$ of the operator $\sigma(F(\lambda)-F_A(\lambda))$ which vanishes at $  \{\lambda_s^T\}$.
We split the set of non-singular points $\lambda_s^T$ in tree subsets $\{\lambda_s^+\}\bigcup\{\lambda_s^-\}\bigcup\{\lambda_s^0\}$, where $\{\lambda_s^+\}$ is the set of non-singular
ITEs for which the corresponding eigenvalues   $\mu_i(\lambda)$ have the following properties: their first nonzero derivative at $\lambda=\lambda_s^+$ has an odd order and it is positive.   If this derivative has an odd order and negative value, then we attribute the corresponding ITE to the set $\{\lambda_s^-\}$, and if it has even order, then   $\lambda_s^T\in\{\lambda_s^0\}$. When an increasing $\lambda$ passes through $\lambda=\lambda_s^+$, the corresponding eigenvalue $\mu_i(\lambda)$ enters the negative semi-axis $(-\infty,0)$ through point $\mu=0$. When $\lambda$ passes through $\lambda=\lambda_s^-$, the corresponding eigenvalue $\mu_i(\lambda)$
exits the negative semi-axis, and $\mu_i(\lambda)$ does not change location relatively to the semi-axis   $(-\infty,0)$ if $\lambda$ passes through $\lambda=\lambda_s^0$.

Denote by $Z^+(\lambda),Z^-(\lambda), Z^0(\lambda)$ the counting functions for the sets $\{\lambda_s^+\},$ $\{\lambda_s^-\},$  and $\{\lambda_s^0\}$, respectively.
For example, $Z^+(\lambda)=\#\{\lambda_s^+<\lambda\}$. Then (see Lemma \ref{ites})
$$
N_T(\lambda) = Z^+(\lambda)+Z^-(\lambda)+Z^0(\lambda)+R(\lambda)\geq Z^+(\lambda)-Z^-(\lambda)+R(\lambda). $$

The change   $n_2=n_2(\lambda') $ in the number of negative eigenvalues $\mu_i(\lambda)$ of the operator   $\sigma(F(\lambda)-F_A(\lambda))$ due to the passage some of the eigenvalues through the origin is equal to $n_2=Z^+(\lambda')-Z^-(\lambda')$.  This and the above estimate for $N_T(\lambda)$ imply that the right-hand side of (\ref{nnn}) does not exceed $N_T(\lambda)$. Thus
 (\ref{nnn}) justifies (\ref{mainIneq}) since $n(\lambda)\geq 0$.

In order to complete the proof of Theorem \ref{lemmamain1}, it remains only to prove lemma \ref{ms}.

{\bf Proof of Lemma \ref{ms}.}
Let us prove (\ref{mainlemma}) in the case when $\lambda_0 \in\{\lambda_i\}$, but $\lambda_0 \notin\{\lambda^A_i\}$. In fact, $m=0$ in this case (see Lemma \ref{ites}). Moreover, $m_A$ is also zero in this case. Thus we need to show that $n_1(\lambda_0)=-\sigma m_0$.

Let $G=G(x,y,\lambda)$ be the Green function of the Dirichlet problem for the Helmgholtz equation (\ref{Anone0}), i.e.,
$$(-\Delta-\lambda)G=\delta(x-y), \quad x,y\in\mathcal O; \quad ~G=0, \quad x\in\partial\mathcal O.
$$
Then
\begin{equation}\label{pred1}
G(x,y,\lambda)=\sum_{n=1}^{\infty} \frac{1}{\lambda_n-\lambda} \Psi^D_n(x) \Psi^D_n(y), \quad x,y \in \mathcal O,
\end{equation}
where $\{\Psi^D_n(x)\}$ is an orthonormal basis of eigenfunctions of the Dirichlet problem for $-\Delta$ in $\mathcal O$, and $\lambda_n$ are the corresponding eigenvalues. Formula (\ref{pred1}) must be understood in the operator sense: for a rigorous meaning, one needs to replace the left-hand side by the operator with the kernel $G$, i.e., by the resolvent $R_\lambda=(-\Delta-\lambda)^{-1}$, and understand the right-hand side as a series of the one-dimensional operators whose kernels are under the summation sign in   (\ref{pred1}). Then (\ref{pred1}) holds as an equality of the operators in $L^2(\mathcal O)$, and
the series in the right-hand side converges in the operator norm.

Furthermore, the solution of the Dirichlet problem for the homogeneous equation (\ref{Anone0}) has the form 
\begin{equation}\label{grgr}
u(x)= -\int_{\partial\mathcal O}\frac{\partial G(x,y,\lambda)}{\partial   \nu_y}u(y)dS_y.
\end{equation}
Purely formally, we substitute (\ref{pred1}) for $G$ in the formula above and then take the normal derivative of both sides at the boundary. This leads to   the following formula for the kernel of the operator $F(\lambda)$: 
$$
F(x,y,\lambda)=\sum_{s=1}^{\infty} \frac{1}{\lambda-\lambda_s} \frac{\partial \Psi^D_s(x)}{
\partial \nu} \frac{\partial \Psi^D_s(y)}{\partial \nu}, \quad x,y \in \partial \mathcal O.
$$
Perhaps this formula does not make sense. However, we are going to show that the corresponding formula for the difference of the operators $F(\lambda)$ and $F(0)$ is valid, i.e., 
\begin{equation}\label{deltaf}
F(\lambda)-F(0)=\sum_{s=1}^{\infty} \frac{\lambda}{\lambda_s(\lambda-\lambda_s)} \frac{\partial \Psi^D_s(x)}{
\partial \nu} \frac{\partial \Psi^D_s(y)}{\partial \nu},
\end{equation}
where the right-hand side is understood as a series of one-dimensional operators   from $H^{3/2}(\partial\mathcal O)$ to $ H^{1/2}(\partial\mathcal O)$ whose kernels are under the summation sign.

In order to justify (\ref{deltaf}), we fix an arbitrary $u_0\in H^{3/2}(\partial\mathcal O)$ and consider the solutions $u,v\in H^{2}(\mathcal O)$ of the equations $(-\Delta-\lambda)u=0$ and $-\Delta v=0$ in $\mathcal O$ with the Dirichlet data $u_0$ at the boundary. Then
\begin{equation}\label{deltaf1}
[F(\lambda)-F(0)]u_0=\frac{\partial (u-v)}{\partial \nu}, \quad x\in \partial\mathcal O.
\end{equation}

From the Green formula it follows that 
\begin{equation}\label{tel1}
\int_\mathcal O \Psi^D_s(y)u(y)dy= \frac{1}{\lambda-\lambda_s}\int_{\partial\mathcal O}\frac{\partial \Psi^D_s(y)}{\partial \nu}u_0(y)dS_y.
\end{equation}
We multiply both sides by $\Psi^D_s(x)$ and sum up the equalities: 
$$
\sum_{s=1}^\infty\int_\mathcal O \Psi^D_s(x)\Psi^D_s(y)u(y)dy= \sum_{s=1}^\infty\int_{\partial\mathcal O}\frac{1}{\lambda-\lambda_s}\Psi^D_s(x)\frac{\partial \Psi^D_s(y)}{\partial \nu}u_0(y)dS_y.
$$
The left-hand side converges to $u(x)$ in $L^2(\mathcal O)$. One can't guarantee a better convergence since $u$ has an inhomogeneous boundary condition.

To proceed with the justification of (\ref{deltaf}), we consider (\ref{tel1}) together with the same formula, where $\lambda=0$ and $u$ is replaced by $v$. From these two formulas it follows that 
$$
w_s := \int_{\mathcal O} \Psi^D_s(x) \Psi^D_s(y) (u(y) - v(y))dy = \frac{\lambda}{\lambda_s(\lambda-\lambda_s)}\int_{\partial\mathcal O}\Psi^D_s(x)\frac{\partial \Psi^D_s(y)}{\partial \nu}u_0(y)dS_y.
$$
The series $\sum w_s$ converges to $u\!-\!v$ in $H^2(\mathcal O)$. Indeed, these series  converges to $u\!-\!v$ in $L_2(\mathcal O)$, and each term vanishes on  $\partial \mathcal O$. Thus in order to justify the convergence in $H^2(\mathcal O)$, it is enough to show that the series $\sum \Delta w_s$ converges in $L_2(\mathcal O)$. The latter follows from the relations
$$
\sum_s \int_{\mathcal O} [\Delta \Psi^D_s(x)] \Psi^D_s(y) (u(y) - v(y))dy = -\sum_s \int_{\mathcal O} \lambda_s \Psi^D_s(x) \Psi^D_s(y) (u(y) - v(y))dy =
$$
$$
 \sum_s \int_{\mathcal O}  \Psi^D_s(x) [\Delta \Psi^D_s(y)] (u(y) - v(y))dy = \sum_s \int_{\mathcal O}  \Psi^D_s(x) \Psi^D_s(y) [\Delta (u(y) - v(y))]dy .
$$
The convergence of $\sum w_s$ in $H^2(\mathcal O)$ and (\ref{deltaf1}) justify (\ref{deltaf}).

When $\lambda$ is in a small neighborhood of $\lambda_0\in \{\lambda_n\}$, the right-hand side in (\ref{deltaf}) is a sum of an analytic in $\lambda$ operator $K(\lambda)$ and a finite-dimensional operator $\frac{1}{\lambda-\lambda_0} P$, where $P$ has the kernel 
$$
P(x,y)= \sum_{s=N+1}^{N+m_0}   \frac{\partial \Psi^D_s(x)}{
\partial \nu} \frac{\partial \Psi^D_s(y)}{\partial \nu}, \quad x,y \in \partial \mathcal O.
$$
Since $P$ is an infinitely smoothing operator, $K(\lambda)$ has the same properties as the operator $F(\lambda)$ , i.e., it is an elliptic pseudo-differential operator with the same principal symbol. Hence
there exist constants $\delta, M$ such that 
\begin{equation}\label{po1}
\sigma(F(\lambda)-F_A(\lambda))=\sigma(K(\lambda)-F_A(\lambda))-\frac{\sigma}{\lambda-\lambda_0} P, \quad
\end{equation}
where (see lemma \ref{lemmaShubin}) 
\begin{equation}\label{po}
\sigma(K(\lambda) - F_A(\lambda))>-M,~~ ~
|\lambda-\lambda_0|<\delta.
\end{equation}
The relations (\ref{po1}), (\ref{po}) will immediately imply that $n_1(\lambda_0)=-
\sigma m_0$ if we show that $P$ is a non-negative $m_0$-dimensional operator. Let us show that $P$ has these properties. In fact, the
set of functions
$\{ \frac{\Psi^D_s}{\partial \nu} \}$ which corresponds to the same Dirichlet eigenvalue   $\lambda_0$ is linearly independent, since if a harmonic function $\Psi$ satisfies the homogeneous Dirichlet and Neumann boundary conditions simultaneously, then $\Psi\equiv 0$. Thus $P$ is $m_0$-dimensional. Furthermore, a one-dimensional symmetric operator of the form   $\varphi (\varphi, \cdot)$ is non-negative. A sum of non-negative operators is also non-negative, i.e., $P$ is non-negative. The lemma  is proved in the case of $\lambda_0\in\{\lambda_i\},~\lambda_0\notin\{\lambda^A_i\}$.

In order to prove the corresponding statement when $\lambda_0\in\{\lambda_i^A\},~\lambda_0\notin\{\lambda_i\}$, consider the Green function $G=G(x,y,\lambda)$ of the Dirichlet problem for equation (\ref{AnoneB}), i.e., the solution of the problem 
\begin{equation}\label{AnoneG}
-\nabla_x A \nabla_x G - \lambda   n(x) G =\delta(x-y), \quad x,y \in \mathcal O \backslash V; \quad
G=0, \quad x,y \in \partial V\bigcup \partial \mathcal O.
\end{equation}
The Green formula for $G$ and a solution $u$ of the homogeneous equation (\ref{AnoneB}) implies
$$
u(x)= \int_{\mathcal O \backslash \mathcal V} [-\nabla_y A \nabla_y G(x,y,\lambda)]u dy -\int_{\mathcal O \backslash \mathcal V} [-\nabla_y A \nabla_y u] G(x,y,\lambda)dy=
$$
$$
-\int_{\partial \mathcal O} \frac{\partial G(x,y,\lambda)}{\partial \nu_{Ay}} u(y) dS_y + \int_{\partial \mathcal V} \frac{\partial u(y)}{\partial \nu_{Ay}}  G(x,y,\lambda) dS_y.
$$
If $u=0$ on $\partial V$, then (\ref{grgr}) is valid. After that, the proof of the statement of the lemma when   $\lambda\in\{\lambda_i^A\},~\lambda\notin\{\lambda_i\}$ is no different from the proof when $\lambda\in\{\lambda_i\}~\lambda\notin\{\lambda_i^A\}$. One needs only to replace $\{ \Psi^D_i\}$ in (\ref{pred1}) by the eigenfunctions $\{ \Psi^A_i\}$ of the problem (\ref{AnoneB}) with the Dirichlet boundary conditions on $\partial O\bigcup\partial V$ and note that the functions   $\{ \frac{\partial \Psi^A_i}{\partial \nu_A} \}$ on   $\partial O$, which correspond to the same eigenvalue of the problem (\ref{AnoneB}), are linearly independent due to the uniqueness of the solution to the Cauchy problem for equation (\ref{AnoneB}).

Consider now the last case:   $\lambda_0\in\{\lambda_i^A\}\bigcap\{\lambda_i\}$. Then, similarly to (\ref{po1}), (\ref{po}), one can show that 
\begin{equation}\label{po1a}
\sigma(F(\lambda)-F_A(\lambda))=L(\lambda)+\frac{\sigma}{\lambda-\lambda_0} P, \quad L(\lambda)>-M,~~|\lambda_0-\lambda|<\delta,
\end{equation}
where the operator $L$ is analytic in $\lambda$ and the kernel of the operator $P$ has the form:
\begin{equation}\label{poa}
P=P_A-P_D= \sum_{s=N}^{N+m_A}   \frac{\partial \Psi^A_s(x)}{
\partial \nu} \frac{\partial \Psi^A_s(y)}{\partial \nu}-\sum_{s=N_1}^{N_1+m_0}   \frac{\partial \Psi^D_s(x)}{
\partial \nu} \frac{\partial \Psi^D_s(y)}{\partial \nu}, \quad x,y \in \partial \mathcal O.
\end{equation}
The summation in the first (second) sum above is over all $s$ which correspond to the eigenfunctions of the Dirichlet problem for equation (\ref{AnoneB}) ((\ref{Anone0}), respectively) with the eigenvalue $\lambda=\lambda_0$. It follows immediately from (\ref{po1a}), (\ref{poa}) that
\begin{equation}\label{nmm}
\delta n_1(\lambda_0)=\sigma(sgn^+ -sgn^-),
\end{equation}
where $sgn^+ (sgn^-)$ is the number of positive (negative) eigenvalues of the operator $ P$.

Let   $V_A$ be the range of the operator $P_A$ (spanned by $\{\frac{\partial \Psi^A_s(x)}{
\partial \nu},~N<s\leq N+m_A\}$, let $V_D$ be the range of the operator $P_D$, and let $V=V_A\bigcap V_D.$ Then dim$V_A=m_A,$ dim$V_D=m_D$, and dim$V=m$. The latter follows from Lemma \ref{ites}. The operator $P$ is positive on functions from $V_A$ which are orthogonal to $V$, and it is negative on
functions from $V_D$ which are orthogonal to $V$. Thus, $m_A-m\leq sgn^+\leq m_A,~m_D-m\leq sgn^-\leq m_D,$ and therefore (after subtraction) $|(sgn^+- sgn^-)-(m_A-m_D)|\leq m$. This and (\ref{nmm}) imply (\ref{mainlemma}).
\qed

\subsection{Proof of Theorem \ref{lemmamain2}}\label{part3}

\textbf{Proof.} From the definition of $\sigma$ it follows that $\sigma=1$ if $A<I$ and $\sigma=-1$ if $A>I$. Let us fix an arbitrary $0\neq v \in H^{3/2}(\partial\mathcal O)$. Denote by $v_0,v_A$ the solutions of the equations  (\ref{Anone0}), (\ref{Anone}), respectively, with the Dirichlet data $v$ at the boundary. If $A>I$  and $v_A$ is extended by zero on $\mathcal V$, then
\begin{equation*}
\int_{\mathcal O} A \nabla v_A \cdot\nabla \overline{v_A }dx>\int_{\mathcal O}  |\nabla v_A|^2 dx\geq\int_{\mathcal O}  |\nabla v_0|^2 dx.
\end{equation*}
The second inequality is an immediate consequence of the definition of  $v_0$ as the function where the corresponding Dirichlet form has the minimum. Similarly, if $A<I$, then
\begin{equation*}
\int_{\mathcal O} A \nabla v_A \cdot\nabla \overline{v_A }dx \leq \int_{\mathcal O} A \nabla v_0 \cdot\nabla \overline{v_0 }dx < \int_{\mathcal O}  |\nabla v_0|^2 dx.
\end{equation*}
Thus in all the cases, 
\begin{equation}\label{energy}
\sigma[\int_{\mathcal O}   |\nabla v_0|^2 dx - \int_{\mathcal O} A \nabla v_A \cdot\nabla \overline{v_A }dx]>0.
\end{equation}

From the Green formula for the equations (\ref{AnoneB}) and (\ref{Anone0}) with $\lambda=0$ it follows that
\[
\int_{\partial\mathcal O} F_A(0)v\cdot\overline{v}dS = \int_{\mathcal O} A \nabla v_A \cdot\nabla \overline{v_A }dx, \quad
\int_{\partial\mathcal O} F(0)v\cdot\overline{v}dS = \int_{\mathcal O}  |\nabla v_0|^2 dx].
\]
This and (\ref{energy}) imply that 
\[
\int_{\partial\mathcal O}\sigma [F(0)-F_A(0)]v\cdot\overline{v}dS>0,
\]
i.e., the operator $\sigma [F(0)-F_A(0)]$ is positive and can not have negative eigenvalues.
\qed

\subsection{Proof of Theorem \ref{ThEstFK}}
 We take into account that $n^-(0)=0$, and note that $\lambda_1^T$ can not exceed any root of the equation
\begin{equation}\label{min}
2=\sigma(N(\lambda)-N_A(\lambda)),~\lambda >0,
\end{equation}
since otherwise the left-hand side in (\ref{mainIneq}) at the root of the equation does not exceed one, while the right-hand side is two.
 Since $N_A(\lambda)=N(\frac{n}{a}\lambda)$,   (\ref{min}) takes the form
$$
2=\sigma(N(\lambda) - N(\frac{n}{a}\lambda)),~ \lambda >0.
$$
Let $\sigma=1$. Since $N(\lambda_2)=2$, it follows that $\lambda_0=\lambda_2$ satisfies (\ref{min}) if $N_A(\lambda_0)=0$, i.e.,
$$
\frac{n}{a} \lambda_2 < \lambda_1.
$$
Let now $\sigma=-1$. Then $N_A(\frac{a}{n}\lambda_2)=2$, and $\lambda_0=\frac{a}{n}\lambda_2$ satisfies  (\ref{min}) if $N(\lambda_0)=0$, i.e.,
$$
\frac{a}{n} \lambda_2 < \lambda_1.
$$
\qed

The remark after Theorem 3.2 can be justified similarly.

\section{Appendix. Ellipticity of the
 porblem and the discreteness of ITEs}
Let us recall conditions on $A,n$, obtained in \cite{LV4}, which guarantee the ellipticity of the interior transmission problem (\ref{Anone0}),(\ref{AnoneB}),(\ref{Antwo}) and the discreteness of its spectrum. Complex-valued $A,n$ are considered in \cite{LV4}, but here we will discuss only the case of real $A$ and $n$.

Let us fix an arbitrary point $x^0\in \partial \mathcal O $ and choose a new orthonormal basis $\{e_j\},~1\leq j\leq d,$ centered at the point $x^0$ with $e_d=\nu$, where $\nu$ is the outer normal to the boundary at the point $x^0$.   The vectors $e_1,...,e_{d-1}$ belong to the tangent plane to $\partial \mathcal O $ at the point $x_0$. Let $y$ be the local coordinates defined by the basis $\{e_j\}$, and let $C=C(x^0)$ be the transfer matrix, i.e., $y=C(x-x^0)$.

We fix the point $x=x^0$ in   equations (\ref{Anone}), (\ref{Antwo}) and rewrite the problem in the local coordinates $y$. Then we get the following problem with constant coefficients in the half space $y_d>0:$
\begin{equation}\label{Anone1}
\begin{array}{l}
-\Delta_y u - \lambda  u =0, \quad y_d>0 ,\\
-\nabla_y \widetilde{A} \nabla_y v - \lambda n(x^0)v =0,   \quad y_d>0,
\end{array}
\end{equation}
\begin{equation}\label{Antwo1}
\begin{array}{l}
u-v=0, \quad y_d=0, \\
\frac{\partial u}{\partial {y_d}} - \frac{\partial v}{\partial \nu_{\widetilde{A}}}=0, \quad y_d=0.
\end{array}
\end{equation}
Here
$$
\widetilde{A}=\widetilde{A}(x^0)=CA(x^0)C^*.
$$
The entries of the matrix $\widetilde{A}=(a_{i,j})$ are equal to $a_{i,j}=e_j\cdot A(x^0)e_i$. The co-normal derivative in the boundary condition equals $e_d\cdot \widetilde{A}\nabla_y.$

The following two theorems were proved by the authors of this paper in \cite{LV4}. Recall that a boundary value problem is elliptic if the equations are elliptic and the Shapiro-Lopatinskii condition holds at the boundary. The latter condition means that, after the Fourier transform with respect to variables $y'=(y_1,... ~y_{d-1})$, the resulting problem on the half line $y_d\geq 0$ has only the trivial stable solution.

\begin{theorem}\label{a3} Let $A(x)>0, n(x)>0$ for $   x \in \overline{\mathcal O}$. Then
 the ellipticity of the problem (\ref{Anone0}),(\ref{AnoneB}),(\ref{Antwo})
is equivalent to the following condition (which is imposed on the matrix $A$ at the boundary of the domain):

 if $d=2$, then
\begin{equation}\label{m3a}
{\rm{det}} A(x^0)\neq 1,\quad   x^0   \in \partial \mathcal O;
\end{equation}

if $d=3$, then
\begin{equation}\label{m3}
det \left ( \begin{array}{ll} a_{3,3}a_{1,1}-(a_{1,3})^2 -1 & a_{3,3}a_{1,2}-a_{1,3}a_{2,3} \\
a_{3,3}a_{2,1}-a_{1,3}a_{2,3} & a_{3,3}a_{2,2}-(a_{2,3})^2 -1 \end{array} \right ) >0, \quad x^0\in \partial \mathcal O.
\end{equation}
\end{theorem}

\begin{theorem}\label{cond1}
If the problem (\ref{Anone0}),(\ref{AnoneB}),(\ref{Antwo}) is elliptic and  additionally
$$
a_{d,d}n(x^0)-1 \neq 0, \quad x^0   \in \partial \mathcal O,
$$
then the spectrum of the interior transmission problem is discrete.
\end{theorem}

The following remarks concern both theorems stated above.

\textbf{Remark 1.} From (\ref{symbols}) it follows that conditions (\ref{m3a}),   (\ref{m3})  are equivalent to the ellipticity of the second order pseudo-differential operator $(F_A)^2-F^2$. Since the operators $(F_A)^2-F^2$ and $(F_A+F)(F_A-F)$ have the same principal symbols, it follows that the ellipticity of the problem (\ref{Anone0})-(\ref{Antwo}) is equivalent to the ellipticity of the operator $F_A(\lambda)-F(\lambda)$.

   Note that the operator $F_A(\lambda)-F(\lambda)$ depends meromorphically on $\lambda$, and satisfies all the properties (to be finitely meromorphic family of Fredholm operators \cite{bleher}) which allow one to make the following conclusion. If this operator is invertible for at least one value of $\lambda_0 \not \in \mathbb C$, then the inverse operator is meromorphic in $\lambda$, and therefore the set of ITEs is at most countable with the only possible accumulation point at infinity.
\\\textbf{Remark 2.} As easy to see, the parameter $\sigma$ introduced in section 2 could be defined differently. If  $d=2$, then
$$
\sigma = sgn(p-p_A)=sgn(1-{\rm{det}} A(x^0)).
$$
If $d=3$, then $\sigma=1$ when the matrix under the determinant sign in (\ref{m3}) is negative, and $\sigma=-1$ when this matrix is positive(this matrix is sign-definite due to (\ref{m3})).

It was also shown in \cite{LV4} that the sign-definiteness of $I\!-\!A$ on $\partial \mathcal O$ is sufficient (but not necessary) for the ellipticity of the interior transmission problem. Moreover, if $I\!-\!A$ is sign-definite on $\partial \mathcal O$, then $\sigma = sgn(p-p_A)=sgn(I-A)|_{\partial \mathcal O}$.

\textbf{Remark 3.} Paper \cite{LV4} concerns the case $\mathcal V=\emptyset$. However, the Shapiro-Lopatinskii conditions and the parameter ellipticity condition (used to prove Theorem \ref{cond1}) must be checked at each part of the boundary independently. Thus problem (\ref{Anone0}),(\ref{AnoneB}),(\ref{Antwo}) is elliptic or parameter-elliptic if and only if the same is true for problem (\ref{Anone0}),(\ref{Anone}),(\ref{Antwo}). The symbol of the operator  $F_A(\lambda)-F(\lambda)$ also does not depend on the presence of an obstacle (the obstacle changes $F_A(\lambda)-F(\lambda)$ by an infinitely smoothing operator). Hence Theorems \ref{a3}, \ref{cond1} and Remarks 1 and 2 remain valid when
$\mathcal V \neq\emptyset$, and their proofs do not require any changes in this more general case.

The condition on $n$ for the discreteness of the spectrum can be weakened if $\mathcal V=\emptyset$ and a stronger requirement is imposed on $A$. The following result is proved in \cite{haddar1}.
\begin{theorem}\label{cond2}
Let $\mathcal V=\emptyset$ and let the matrix $A-I$ be sign-definite for all $x   \in   \overline{\mathcal O}.$  If
$$
\int_{\mathcal O} n(x) dx \neq 1,
$$
then the spectrum of the interior transmission problem is discrete.
\end{theorem}


\begin{thebibliography}{102}

\bibitem{bleher} Bleher, P. M. Operators that depend meromorphically on a parameter. (Russian) Vestnik Moskov. Univ. Ser. I Mat. Meh. 24 1969 no. 5 30-36.

\bibitem{BK}
K. Kh. Boimatov, A. G. Kostyuchenko, Spectral asymptotics of nonselfadjoint elliptic systems of differential operators on bounded domains, Mat. Sb., 181:12 (1990), 1678-1693.
\bibitem{haddar1}      A.-S. Bonnet-BenDhia, L. Chesnel and H. Haddar, {\it On the use of T-coercivity to study the Interior Transmission Eigenvalue Problem},     C. R. Acad. Sci., Ser. I, vol. 340, 2011.


\bibitem{chobst}
F. Cakoni, A. Cossonniere and H. Haddar, Transmission eigenvalues for inhomogeneous media containing obstacles, Inverse problems and imaging.

\bibitem{HadCak} F. Cakoni, H. Haddar, Transmission Eigenvalues in Inverse Scattering Theory, 2012.

\bibitem{Dr} F. Cakoni, D. Gintides, and H. Haddar. The existence of an infinite discrete set of
transmission eigenvalues. SIAM J. Math. Anal., 42:237-255, 2010.

\bibitem{CK}F. Cakoni and A. Kirsch. On the interior transmission eigenvalue problem, Int. Jour.
Comp. Sci. Math., 3:142-167, 2010.

\bibitem{EP2}
J.P.Eckmann, C.-A. Pillet, Zeta functions with Dirichlet and Neumann boundary conditions for exterior domains, Helv. Phys. Acta, 70, 44-65, 1997.


\bibitem{KS} Yu.Egorov, A.Komech,M.Shubin, Elements of the Modern Theory of Partial Differential Equations, Springer, 2001.

\bibitem{fried}
L. Friedlander, Some inequalities between Dirichlet and Neumann eigenvalues, Archive for Rational Mechanics and Analysis, 116, 153--160, 1991.



\bibitem{tsm}
M.Hitrik, K.Krupchyk, P.Ola, L.Paivarinta, Transmission eigenvalues for elliptic operators, SIAM J. Math. Anal., 43, 2630-2639, 2011.

\bibitem{LV4}
E.Lakshtanov, B.Vainberg, Ellipticity in the interior transmission problem in anisotropic media, SIAM J. Math. Anal. 44, pp. 1165-1174, 2012.

\bibitem{lakvain5}
E.Lakshtanov, B.Vainberg, Remarks on interior transmission eigenvalues, Weyl formula and branching billiards, J. Phys. A: Math. Theor. 45, 125202, 2012.


\bibitem{reedsimon} M. Reed, B. Simon, Methods of Modern Mathematical Physics, IV, Academic Press,
1978.

\bibitem{safvas}
Yu. Safarov and D. Vassiliev, The Asymptotic Distribution of Eigenvalues of Partial Differential Operators, American Mathematical Society, (1997, 1998).



\bibitem{saf} Yu. Safarov, On the comparison of the Dirichlet and Neumann counting functions. AMS Translations (2), Advances in Mathematical Sciences, 225,191-204, 2008.


\bibitem{ss}
V.Serov, J.Sylvester, Transmission Eigenvalues: some degenerate and singular cases, Inverse Problems, 28, 065004, 2012.

\bibitem{shubin} M.Shubin, Pseudodifferential Operators and Spectral Theory, Springer, 2001.



\bibitem{yaf}D. R. Yafaev, The Schrodinger Operator: Perturbation Determinants, the Spectral Shift Function, Trace Identities, and All That, Funkts. Anal. Prilozh., 41:3, 60-83, 2007.


\end{thebibliography}
\end{document}